\documentclass[published]{JHEP3} 
\JHEP{00(2008)000}
\JHEPspecialurl{http://jhep.sissa.it/JOURNAL/JHEP3.tar.gz}
\usepackage{amssymb,epsfig}

\newcommand\fverb{\setbox\fverbbox=\hbox\bgroup\verb}
\newcommand\fverbdo{\egroup\medskip\noindent%
                        \fbox{\unhbox\fverbbox}\ }
\newcommand\fverbit{\egroup\item[\fbox{\unhbox\fverbbox}]}
\newbox\fverbbox

\def\be{\begin{eqnarray}}
\def\ee{\end{eqnarray}}
\def\bec{\begin{center}}
\def\ec{\end{center}}

\def\beq{\begin{equation}}
\def\eeq{\end{equation}}
\def\haf{\frac{1}{2}}

\def\f{\frac}

\title{Singlet fermionic dark matter}

\author{Kang Young Lee \\
        Department of Physics, Korea University, Seoul 136-701, Korea \\
        E-mail: \email{kylee@muon.kaist.ac.kr}}

\author{Yeong Gyun Kim,~ Seodong Shin\\
        Department of Physics, KAIST, Daejeon 305-701, Korea \\
        E-mail: \email{ygkim@muon.kaist.ac.kr},
               ~\email{sshin@muon.kaist.ac.kr} }

\received{} 
\revised{}
\accepted{}           

\preprint{ KAIST-TH 2007/10 }     

\abstract{
We propose a renormalizable model of a fermionic dark matter 
by introducing a gauge singlet Dirac fermion
and a real singlet scalar.
The bridges between the singlet sector and the standard model sector
are only the singlet scalar interaction terms with
the standard model Higgs field.
The singlet fermion couples to the standard model particles
through the mixing between the standard model Higgs and singlet scalar 
and is naturally a weakly interacting massive particle (WIMP).
The measured relic abundance can be explained 
by the singlet fermionic dark matter as the WIMP within this model. 
Collider implication of the singlet fermionic dark matter is also discussed.
Predicted is the elastic scattering cross section of the singlet fermion
into target nuclei for a direct detection of the dark matter.
Search of the direct detection of the dark matter 
provides severe constraints on the parameters of our model.
 }


\dedicated{}

\begin{document}


\section{Introduction}

The missing mass of some non-visible form of matter 
in the galaxy cluster was first investigated 
by Zwicky in 1933 \cite{Zwicky}. 
Since then, there have been a lot of efforts
to probe the dark matter (DM) which provides the unseen mass in the cluster. 
Evidences have been found including 
the galactic rotation curve \cite{rotation} 
and the observation of the Bullet cluster \cite{Bullet}. 
The precise measurement of the relic abundance 
of the cold dark matter (CDM) has been obtained from the
Wilkinson microwave anisotropy probe (WMAP) data on the cosmic
microwave background radiation as \cite{WMAP}
\be
0.085 < \Omega_{CDM} h^2 < 0.119, ~~~~ (2\sigma~\mbox{level})
\ee
where $\Omega$ is the normalized relic density and 
the scaled Hubble constant $h \approx 0.7$ 
in the units of 100 km/sec/Mpc.

Since there is no proper CDM candidate in the standard model (SM) contents,
the extended models of the SM are required to provide a candidate of DM.
Various candidates of the CDM have been proposed.
Weakly interacting massive particles (WIMP) 
are favored to explain the observed value of the relic abundance 
in view of new physics beyond the SM. 
WIMPs include the lightest supersymmetric particle (LSP) 
in the supersymmetric models with $R$ parity \cite{LSP1, LSP2}, 
the lightest Kaluza-Klein particle in the extra dimensional models 
with conserved KK parity \cite{LKKP}, 
and the lightest T-odd particle in the T-parity conserved little Higgs model
\cite{Todd}.
Addition of a real singlet scalar field to the SM with $Z_2$-parity
has been considered as one of the simplest extensions of the SM 
with the nonbaryonic CDM \cite{SS1,SS2,SS3}. 
A model which introduces singlet Majorana neutrinos and a singlet scalar
has been considered in Ref. \cite{Y}.
Another scalar extension with multiple Higgs doublets is
presented in Ref. \cite{doublet}.
General classification of the extra gauge multiplets as a minimal dark
matter candidate has been performed in Ref. \cite{CFS}. 
On the other hand, incorporating the gauge coupling unification 
with the dark matter issue, 
a fermionic DM with the quantum numbers of SUSY higgsinos and a
singlet is suggested \cite{Higgsino,DEramo}. Heavy particle DM communicating
with supersymmetric SM via pure Higgs sector interaction, which is motivated
by Higgs portal and Hidden valley models is also considered\cite{Oxford}.
A model with a gauge singlet Dirac fermion is
proposed as a minimal model of fermionic dark matter\cite{KL}.  
In this model, the singlet fermion interacts with the SM sector only 
through nonrenormalizable interactions
among which the leading interaction term is given by the dimension five term 
$(1/\Lambda) H^{\dagger} H \bar{\psi}\psi$,  
where $H$ is the SM Higgs doublet and
$\psi$ is the dark matter fermion,
suppressed by a new physics scale $\Lambda$. 

We propose a renormalizable extension of the SM 
with a hidden sector which consists of SM gauge singlets 
(a singlet scalar and a singlet Dirac fermion).
The singlet scalar interacts with the SM sector 
through the triple and quartic scalar interactions.
There are no renormalizable interaction terms 
between the singlet fermion and the SM particles 
but the singlet fermion interacts with the SM matters 
only via the singlet scalar. 
Therefore it is natural that the singlet fermion is a WIMP
and a candidate of the CDM. 
Our model is a minimal model of renormalizable extension of the SM
including the fermionic dark matter.
The model is described in section \ref{sec:Model}.
We show that the singlet fermion can be a CDM candidate,
which explains the measured relic density by the WMAP with 
the experimental constraint on the Higgs bosons at LEP2
in section \ref{sec:Cosmological implication}. 
The direct detection of the fermionic CDM is investigated 
in Sec.\ref{sec:Direct detection}. 
Finally we conclude in Sec.\ref{sec:Conclusion}.

\section{The model}
\label{sec:Model}
We introduce a hidden sector consisting of a real scalar field $S$ 
and a Dirac fermion field $\psi$ which are SM gauge singlets.
The singlet scalar $S$ couples to the SM particles
only through triple and quartic terms with the SM Higgs boson
such as $S H^\dagger H$ and $S^2 H^\dagger H$.
New fermion number of the singlet fermion is required to be conserved
in order to avoid the mixing between the singlet fermion and the
SM fermions. The global $U(1)$ charge of the singlet Dirac
fermion takes the role of the new fermion number. As a result,  
no renormalizable interaction terms between the singlet fermion $\psi$ 
and the SM particles are allowed.
Thus the interaction of $\psi$ with the SM particles 
just comes via the singlet scalar.

We write the Lagrangian as
\be
\mathcal{L} = \mathcal{L}_{SM} + \mathcal{L}_{hid} + \mathcal{L}_{int}, 
\ee
where the hidden sector Lagrangian is given by
\be
\mathcal{L}_{hid} = \mathcal{L}_{S} + \mathcal{L}_{\psi} 
                       -g_S \bar{\psi}\psi S ,
\ee
with
\be
\mathcal{L}_{S} &=& 
\haf \left(\partial_{\mu} S\right)\left(\partial^{\mu} S\right)
      -\frac{m_0^2}{2} S^2 -\frac{\lambda_3}{3!}S^3-\frac{\lambda_4}{4!}S^4 ,
\nonumber \\
\mathcal{L}_{\psi} &=& \bar{\psi}\left(i\partial\!\!\!/ 
                                 - m_{\psi_0}\right)\psi. \label{eq:ps}
\ee
The interaction Lagrangian between the hidden sector and the SM fields
is given by
\be
\mathcal{L}_{int} = -\lambda_1 H^{\dagger} H S 
                         - \lambda_2 H^{\dagger} H S^2.  \label{eq:int}
\ee
The scalar potential given in Eq. (\ref{eq:ps}) and (\ref{eq:int}) 
together with the SM Higgs potential
$-\mu^2 H^{\dagger} H + \bar{\lambda}_0 (H^{\dagger} H)^2$
derives the vacuum expectation values (VEVs) 
\be
\langle H \rangle = \frac{1}{\sqrt{2}} 
                     \left( \begin{array}{c}
                            0 \\
                            v_0 \\
                            \end{array}
                     \right)
\ee
for the SM Higgs doublet
to give rise to the electroweak symmetry breaking, and
$\langle S \rangle = x_0$ for the singlet scalar sector.
The extremum conditions $\partial V / \partial H |_{<H^0>=v_0/\sqrt{2}}= 0$ and
$\partial V / \partial S |_{<S>=x_0} =0$ 
lead to the relations \cite{PMS}
\be
\mu^2 &=& \bar{\lambda}_0 v_0^2 + (\lambda_1 + \lambda_2 x_0)x_0 ,
\nonumber \\
m_0^2 &=& - \frac{\lambda_3}{2}x_0 -\frac{\lambda_4}{6} x_0^2 
          - \frac{\lambda_1 v_0^2}{2 x_0} - \lambda_2 v_0^2 .
\ee
The neutral scalar states $h$ and $s$ defined by
$H^0=(v_0+h)/\sqrt{2}$ and $S=x_0+s$ are mixed 
to yield the mass matrix given by
\be
\mu_{h}^2 &\equiv& \left.\frac{\partial^2 V}{\partial h^2}\right|_{h=s=0} 
           = 2 \bar{\lambda}_0 v_0^2 ,
\nonumber \\
\mu_{s}^2 &\equiv& \left.\frac{\partial^2 V}{\partial s^2}\right|_{h=s=0} 
           = \frac{\lambda_3}{2} x_0 + \frac{\lambda_4}{3} x_0^2 
                  - \frac{\lambda_1 v_0^2}{2 x_0} ,
\nonumber \\
\mu_{hs}^2 &\equiv& \left.\frac{\partial^2 V}
                               {\partial h \partial s}\right|_{h=s=0} 
           = (\lambda_1 + 2 \lambda_2 x_0) v_0.
\ee
The mass eigenstates $h_1$ and $h_2$ are obtained  by
\be
h_1 &=& \sin \theta \ s + \cos \theta \ h ,
\nonumber \\
h_2 &=& \cos \theta \ s - \sin \theta \ h ,
\ee
where the mixing angle $\theta$ is defined by
\be
\tan \theta = \frac{y}{1+\sqrt{1+y^2}}, 
\ee
with $ \ y \equiv  2 \mu_{hs}^2 /(\mu_h^2 - \mu_s^2)$.
The Higgs boson masses $m_1$ and $m_2$ are given by 
\be
m^2_{1,2} = \frac{\mu_h^2+\mu_s^2}{2} 
            \pm \frac{\mu_h^2-\mu_s^2}{2}\sqrt{1+y^2},
\ee
where the upper (lower) sign corresponds to $m_1(m_2)$.
According to the definition of $\tan \theta$, 
we get $|\cos \theta| > \f{1}{\sqrt{2}}$
implying that $h_1$ is SM Higgs-like 
while $h_2$ is the singlet-like scalars. 
As a result, there exist two neutral Higgs bosons in our model
and the collider phenomenology of the Higgs sector might be affected.
We will discuss it in later section.

The singlet fermion $\psi$ has the mass $m_\psi = m_{\psi_0} + g_S x_0$
as an independent parameter of the model 
since $m_{\psi_0} $ is just a free parameter.
The Yukawa coupling $g_S$ measures the interaction
of $\psi$ with other particles.
Generically the interactions between $\psi$ and the SM particles
are suppressed by the mass of singlet scalar and/or the Higgs mixing.
Therefore $\psi$ is naturally weakly interacting and
can play the role of a cold dark matter as an WIMP.
If we fix masses of two Higgs bosons, 
the singlet fermion annihilation processes into the SM particles 
depend upon the fermion mass $m_\psi$, Yukawa coupling $g_S$, 
and the Higgs mixing angle $\theta$. 
If the final state includes Higgs bosons, $h_1$ or $h_2$,
several Higgs self-couplings are involved
depending on various couplings in the scalar potential.

\section{Implications in cosmology and collider physics}
\label{sec:Cosmological implication}

In the early universe, the dark matter is assumed to be 
in thermal equilibrium by the active annihilation and production process
with the SM sector.
When the universe cools down and the temperature $T$ drops below the DM mass, 
the DM number density is suppressed exponentially 
so that the annihilation rate of the dark matter becomes smaller 
than even the Hubble parameter. 
Then the interactions of the DM freeze out 
to make the DM particles fall out of equilibrium 
and the DM number density in a comoving volume remains constant. 
Therefore the current relic abundance of the CDM depends on 
the annihilation cross section of $\psi$ into the SM particles or
the additional Higgs bosons in our model.
The pair annihilation process of $\psi$ consists of the annihilation
into SM particles via Higgs-mediated $s$-channel processes 
and into Higgs bosons via $s$, $t$, and $u$-channels.
The dominant final states of the SM particles are
$b \bar{b}$, $t \bar{t}$, $W^+ W^-$, and $ZZ$.
The total cross section of the annihilation process is given by
\be
\sigma v_{rel} &=& \frac{(g_S \sin \theta \cos \theta)^2}
                        {16\pi}
                   \left(1-\frac{4 m_{\psi}^2}{\mathfrak{s}}\right) 
\\
      & & \times \left(\frac{1}
            {(\mathfrak{s}-m_{h_1}^2)^2 + m_{h_1}^2 \Gamma_{h_1}^2}
            +\frac{1}
               {(\mathfrak{s}-m_{h_2}^2)^2 + m_{h_2}^2 \Gamma_{h_2}^2}\right.
\\
      & & ~~~~~
      \left.-\frac{2(\mathfrak{s}-m_{h_1}^2)(\mathfrak{s}-m_{h_2}^2) 
                    + 2 m_{h_1}m_{h_2}\Gamma_{h_1}\Gamma_{h_2}}
                  {((\mathfrak{s}-m_{h_1}^2)^2 + m_{h_1}^2 \Gamma_{h_1}^2)
                   ((\mathfrak{s}-m_{h_2}^2)^2 + m_{h_2}^2 \Gamma_{h_2}^2)}
      \right) 
\\
      & & \times \left[
           \left(\frac{m_b}{v_0} \right)^2 \cdot 2 \mathfrak{s} 
                 \left(1-\frac{4 m_b^2}{\mathfrak{s}}\right)^{3/2}\cdot 3 
            + \left(\frac{m_t}{v_0} \right)^2 \cdot 2 \mathfrak{s} 
              \left(1-\frac{4 m_t^2}{\mathfrak{s}}\right)^{3/2}\cdot 3 
      \right.
\\
      & & ~~~~~ 
           + \left(2\frac{m_W^2}{v_0}\right)^2 
             \left(2+\frac{(\mathfrak{s}-2m_W^2)^2}{4 m_W^4}\right)
                 \cdot\sqrt{1-\frac{4 m_W^2}{\mathfrak{s}}} 
\\
      & & ~~~~~ 
    \left. + \left(2\frac{m_Z^2}{v_0}\right)^2
             \left(2+\frac{(\mathfrak{s}-2m_Z^2)^2}{4 m_Z^4}\right)
                 \cdot\sqrt{1-\frac{4 m_Z^2}{\mathfrak{s}}} \cdot \frac12 
         \right]
\\
      & &  + \sum_{i,j=1,2} \sigma_{h_i h_j} 
           + \sum_{i,j,k=1,2} \sigma_{h_i h_j h_k} ,
\ee
where $\Gamma_{h_i}$ is the decay width of $h_i$ for $i=1,2$, 
and $\sigma_{h_i h_j}$, $\sigma_{h_i h_j h_k}$ are 
the annihilation cross sections of $\bar{\psi}\psi$ into $h_i h_j$ 
or $h_i h_j h_k$ with $i,j,k=1,2$. Here, $\sqrt{\mathfrak{s}}$ denotes the center of mass energy. 
The thermal average of the cross section over $\mathfrak{s}$ is given by 
\be
\langle \sigma_{ann.} v_{rel} \rangle 
    = \frac{1}{8 m_{\psi}^4 T K_2^2(m_{\psi}/T)} 
      \int_{4m_{\psi}^2}^{\infty} d\mathfrak{s}~ 
           \sigma_{ann.}(\mathfrak{s}) \left(\mathfrak{s}-4m_{\psi}^2\right)
                \sqrt{\mathfrak{s}} K_{1} \left(\frac{\sqrt{\mathfrak{s}}}{T}\right),
\ee
where $K_{1,2}$ are the modified Bessel functions. 
The evolution of the number density of the singlet fermion is described 
by the Boltzmann equation in terms of 
$ \langle \sigma_{ann.} v_{rel} \rangle$ 
and the equilibrium number density of $\psi$
\be
\frac{d n_{\psi}}{d t} + 3 H n_{\psi} = 
            - \langle \sigma_{ann.} v_{rel} \rangle 
           \left[ n_{\psi}^2 - \left(n_{\psi}^{EQ}\right)^2 \right],
\ee
where $H$ is the Hubble parameter and $n_{\psi}^{EQ}$ is the
equilibrium number density of $\psi$. 
After the freeze out of the annihilation processes, 
the actual number of $\psi$ per comoving volume becomes constant 
and the present relic density $\rho_\psi = m_{\psi} n_{\psi}$ is determined.
The freeze-out condition gives the thermal relic density in terms
of the thermal average of the annihilation cross section.
\be
\Omega_{\psi} h^2 \approx \frac{(1.07 \times 10^9)x_F}
            {\sqrt{g_{*}} M_{pl}(GeV) \langle \sigma_{ann.} v_{rel} \rangle},
\ee
where $g_{*}$ counts the effective degrees of freedom 
of the relativistic quantities in equilibrium. 
The inverse freeze-out temperature $x_F=m_{\psi}/T_F$ 
is determined by the iterative equation
\be
x_F = \log \left( \frac{m_{\psi}}{2\pi^3} 
                  \sqrt{\frac{45 M_{pl}^2}{2 g_{*} x_F}} 
                  \langle \sigma_{ann.} v_{rel} \rangle \right) .
\ee
\FIGURE{
\epsfig{figure=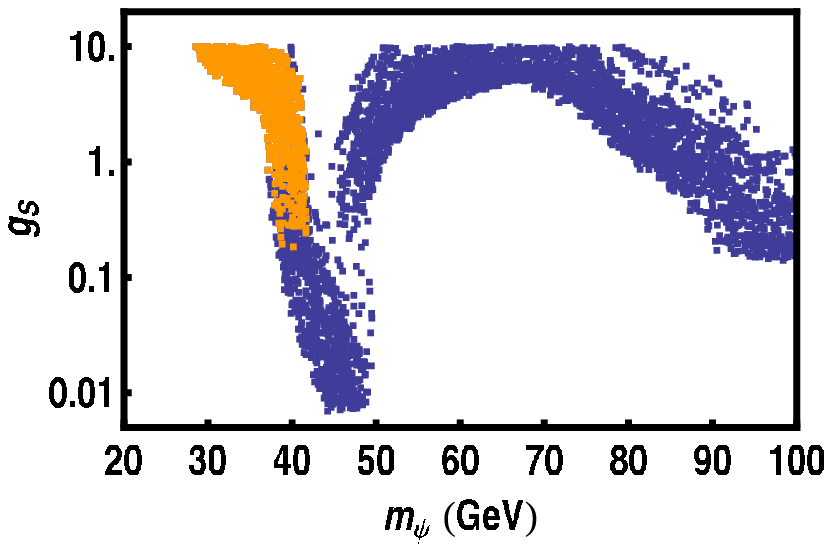,width=10cm,height=8cm}
\caption{ 
Allowed parameter set of $(m_{\psi},g_S)$ 
with $m_{h_1}=90$ GeV $(\pm 1\%)$ and $m_{h_2}=500$ GeV $(\pm 12\%)$. The allowed region by LEP2 data is denoted as orange region. 
}
\label{fig:new90_c}
}
\FIGURE{
\epsfig{figure=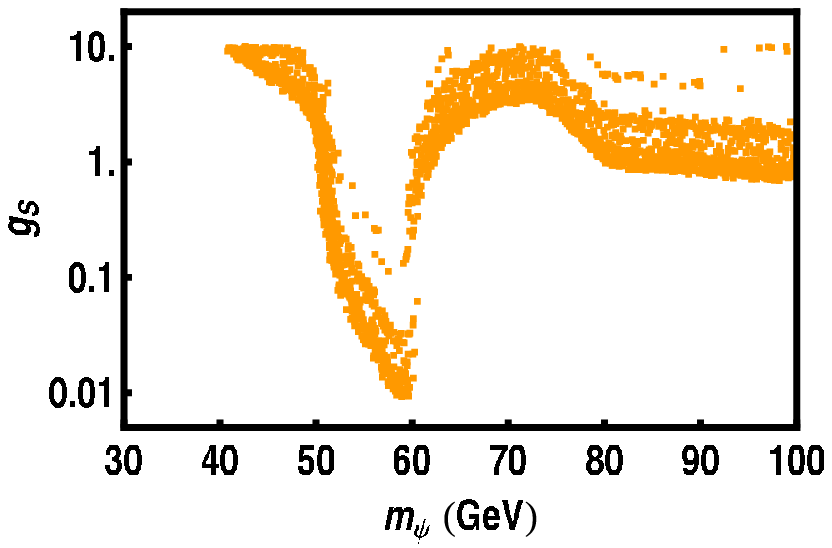,width=10cm,height=8cm}
\caption{ 
Allowed parameter set of $(m_{\psi},g_S)$ 
with $m_{h_1}=120$ GeV $(\pm 1\%)$ and $m_{h_2}=500$ GeV $(\pm 12\%)$. The allowed region by LEP2 data is denoted as orange region. 
}
\label{fig:random120_c}
}
\FIGURE{
\epsfig{figure=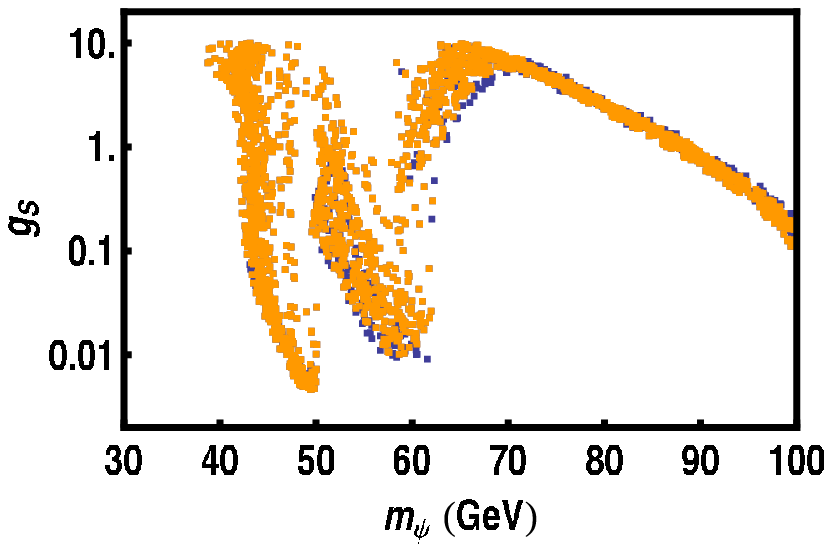,width=10cm,height=8cm}
\caption{
Allowed parameter set of $(m_{\psi},g_S)$ 
with $m_{h_1}=120$ GeV $(\pm 4\%)$ and $m_{h_2}=100$ GeV $(\pm 1\%)$. The allowed region by LEP2 data is denoted as orange region. 
} 
\label{fig:random_mix_c}
}

We investigate the allowed model parameter space, which provide thermal relic density 
consistent with the WMAP observation.
In addition to $m_{\psi}$ and $g_S$, 
we have six more undetermined parameters in the scalar potential:
$\bar{\lambda}_0$, $\lambda_1$, $\lambda_2$, $\lambda_3$, $\lambda_4$, $x_0$, 
which determine the Higgs boson masses ($m_{h_1}$, $m_{h_2}$), mixing angle ($\theta$), 
and triple and quartic self couplings of Higgs bosons.
Here we have to study a large volume of the multidimensional parameter space.
For clarity of the presentation of our result, 
we fix the Higgs masses, $m_{h_1}$ and $m_{h_2}$
within some ranges, while allowing the other parameters such as Higgs mixing angle
and self couplings vary freely.
Our parameter sets should satisfy several physical conditions.
We demand that 
$i)$ the potential is bounded from below,
$ii)$ the electroweak symmetry breaking is viable, 
and $iii)$ all couplings keep the perturbativity. 

Fig.\ref{fig:new90_c} shows the allowed parameter set
by the measured relic abundance with $m_{h_1}=90$ GeV $(\pm 1\%)$
and $m_{h_2}=500$ GeV $(\pm 12\%)$. 
The valley in Fig.\ref{fig:new90_c} implies the resonant region 
of $h_1$ exchange for DM pair annihilation, where $2m_{\psi} \simeq m_{h_1}$. 
The coupling constant $g_S$ should be small in that region 
in order to compensate the enhancement of the cross section 
due to the Higgs resonance effect. 
A step appears when $m_{\psi}\sim 80 GeV$, which denotes that the
annihilation channel $\bar{\psi} \psi \rightarrow W^{-} W^{+} / Z Z$ 
opens as $m_{\psi}$ exceeds the $W$ and $Z$ boson masses. 

The current experimental bound on Higgs mass can have a significant impact 
on the allowed parameter space. 
Note that the LEP2 bound of the SM Higgs boson mass gets weaker in our model
since the SM-like Higgs couplings are modified, 
and therefore $m_{h_1}=90$ GeV might be allowed depending on the parameter set.
The promising channel to produce a neutral Higgs boson at LEP
is the Higgs-strahlung process, $e^- e^+ \to Z h$.
A lower bound on the mass of the SM Higgs boson has been
established to be 114.4 GeV at 95 $\%$ confidence level \cite{LEP2}.
In our model, the Higgs mixing alters the $h_i ZZ$ couplings
and therefore the cross sections of the Higgs-strahlung processes.
Furthermore, Higgses can invisibly decay to a pair of singlet fermions.
Thus the SM bound should be modified accordingly.
We consider the parameters defined by
\be
\xi_i^2 
        = \left( \frac{g_{h_iZZ}}{g_{HZZ}^{SM}} \right)^2 
                 \frac{\Gamma_{h_i}^{SM}}
                   {\Gamma_{h_i}^{SM} + \Gamma(h_i \to \bar{\psi}\psi)} ,
\ee
where $\Gamma_{h_i}^{SM}$ are the widths of $h_i$ 
decays into the SM particles. 
Assuming the non-standard models,
the lower bound on the Higgs mass is represented 
by the upper bound of $\xi_i^2$, which is shown in Ref. \cite{LEP2}. 
In our analysis, we impose $\xi^2 < 0.1$ as a conservative bound for $m_{h_i} = 90$ GeV 
(and $\xi^2 < 0.3$ for $m_{h_i} = 100$ GeV for later use). 
Since a new fermion exists in our model,
the definition of $\xi^2$ includes the decay width for invisible decay channel, 
$h_i \to \bar{\psi}\psi $.
On the figure, the orange points denote the region which satisfy the Higgs mass bound.
The allowed region appears when $m_{\psi} \lesssim m_{h_1}/2$,
due to large contribution from the invisible Higgs decay.
If $m_{\psi}>m_{h_1}/2$, the decay channel $h_1\to\bar{\psi}\psi$ is closed and 
$\xi_1^2=(\cos\theta)^2$ is always larger than $0.1$ so that 
the corresponding region is excluded by the experimental results of LEP2.

Allowed parameter set with $m_{h_1}=120$ GeV and $m_{h_2}=500$ GeV 
are shown in Fig.\ref{fig:random120_c}.
This parameter set always satisfy the LEP2 constraints
on the Higgs boson mass, because the Higgs masses are lager than 
the current experimental bound.

Fig.\ref{fig:random_mix_c} shows the allowed parameter set 
for $m_{h_1} = 120$ GeV and $m_{h_2}=100$ GeV.
Here $m_{h_1}$ and $m_{h_2}$ are comparable and 
there are two resonant regions corresponding to
$h_1$ and $h_2$ resonances. One can notice that the most of parameter points
also satisfy the Higgs mass bound.

\FIGURE{
\epsfig{figure=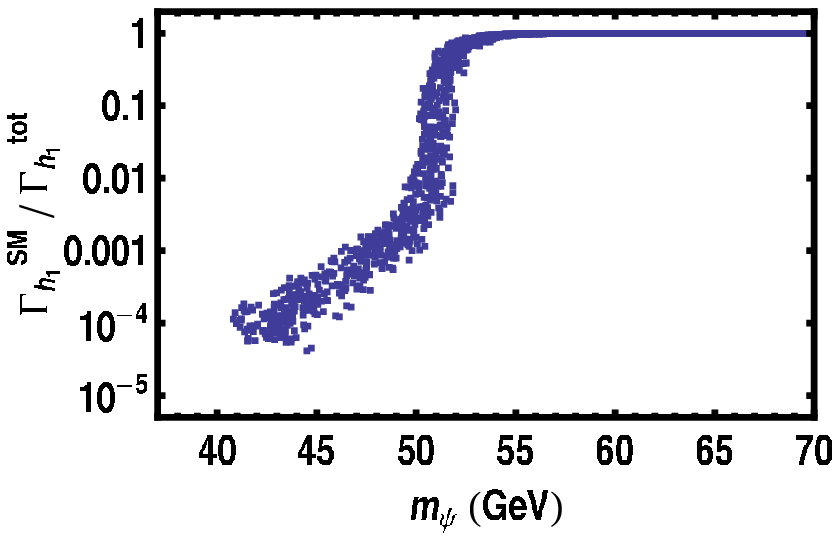,width=10cm,height=8cm}
\caption{
The ratio of $\Gamma_{h_1}^{SM}$ to the total decay width of $h_1$  
with $m_{h_1}=120$ GeV $(\pm 1\%)$ and $m_{h_2}=500$ GeV $(\pm 12\%)$.  
} 
\label{fig:ratio_120}
}

We now briefly comment on possible LHC phenomenology of our model.
One obvious difference of the model from SM is that we have two neutral
Higgs bosons which have smaller couplings to SM particles, compared to SM Higgs case.
Other characteristic of the model is that the Higgs bosons can decay invisibly to
a pair of singlet fermions, if kinematically allowed.
Fig. \ref{fig:ratio_120} shows the ratio of
$\Gamma_{h_1}^{SM}$ to the total decay width of $h_1$, 
for $m_{h_1}=120$ GeV and $m_{h_2}=500$ GeV. 
We have very large invisible branching ratios when the mass of singlet fermion
is less than half of $m_{h_1}$. 
Such a large invisible Higgs decay may be observed 
at the CERN LHC \cite{gunion,eboli}.

\section{Direct detection}
\label{sec:Direct detection}

There are several experiments to detect the WIMP directly 
through the elastic scattering of the WIMP on the target nuclei
\cite{GW,Munoz}.
The effective Lagrangian describing the
elastic scattering of the WIMP and a nucleon is given by
\be
\mathcal{L}_{eff} = f_p (\bar{\psi} \psi)(p \bar{p}) 
                  + f_n (\bar{\psi}\psi) (n \bar{n}) , 
\label{effl}
\ee
where the coupling constant $f_p$ is given by \cite{NS,Ellis}
\be
\frac{f_{p,n}}{m_{p,n}}=\sum_{q=u,d,s} f^{(p,n)}_{Tq} \frac{\alpha_q}{m_q} 
          + \frac{2}{27} f^{(p,n)}_{Tg} \sum_{q=c,b,t}\frac{\alpha_q}{m_q},
\ee
with the matrix elements 
$m_{(p,n)} f^{(p,n)}_{Tq} \equiv \langle p,n|m_q \bar{q}q|p,n \rangle$ 
for $q=u,d,s$ and $f^{(p,n)}_{Tg} = 1-\sum_{q=u,d,s} f^{(p,n)}_{Tq}$. 
The numerical values of the hadronic matrix elements 
$f^{(p,n)}_{Tq}$ are determined in Ref. \cite{Ellis}
\be
f^{(p)}_{Tu}=0.020 \pm 0.004, \ \ f^{(p)}_{Td}=0.026 \pm 0.005, \
\ f^{(p)}_{Ts} = 0.118 \pm 0.062 ,
\ee
and
\be
f^{(n)}_{Tu}=0.014 \pm 0.003, \ \ f^{(n)}_{Td}=0.036 \pm 0.008, \
\ f^{(n)}_{Ts} = 0.118 \pm 0.062 .
\ee
\FIGURE{
\epsfig{figure=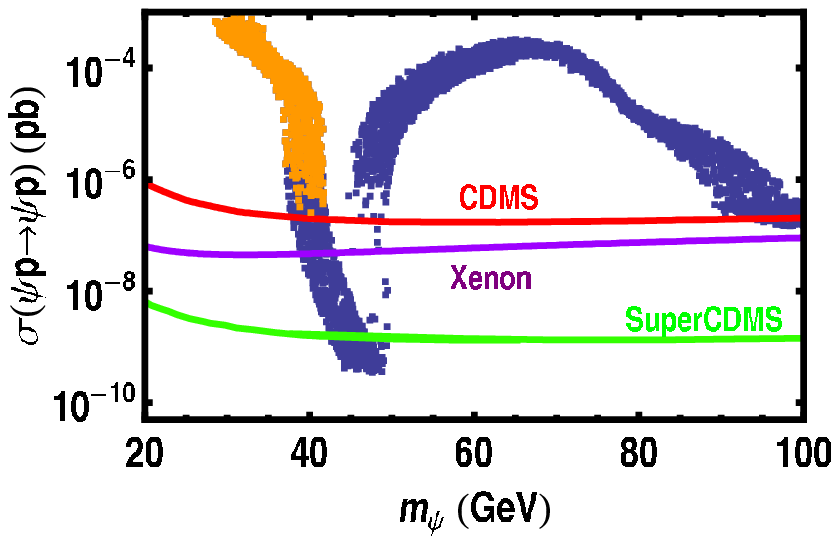,width=10cm,height=8cm}
\caption{
Predictions of the elastic scattering cross section 
$\sigma(\psi p \to \psi p)$ with respect to $m_{\psi}$ 
with $m_{h_1}=90$ GeV $(\pm 1\%)$ and $m_{h_2}=500$ GeV $(\pm 12\%)$. 
The red line indicates the CDMS bound, 
the purple line the Xenon bound, 
and the green line the up-coming super CDMS bound.
The allowed region by LEP2 data is denoted as orange region.
}
\label{fig:new90_j}
}

\FIGURE{
\epsfig{figure=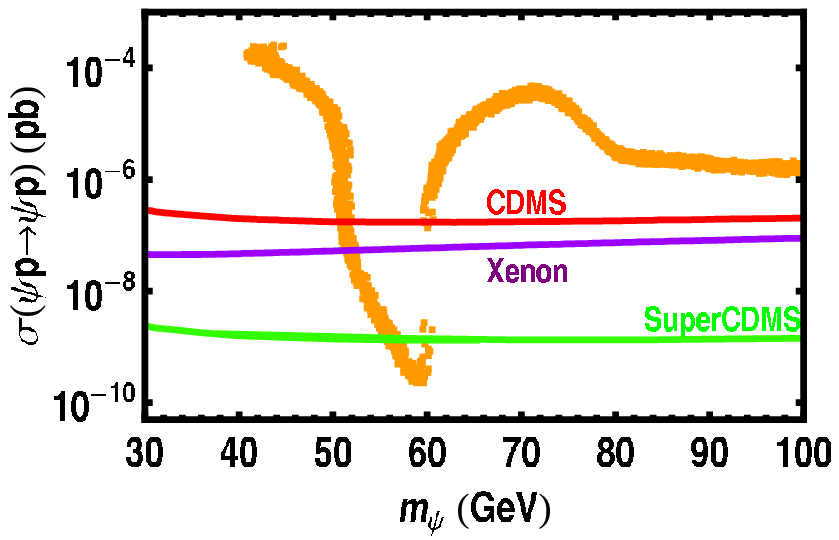,width=10cm,height=8cm}
\caption{
Predictions of the elastic scattering cross section 
$\sigma(\psi p \to \psi p)$ with respect to $m_{\psi}$ 
with $m_{h_1}=120$ GeV $(\pm 1\%)$ and $m_{h_2}=500$ GeV $(\pm 12\%)$. 
The red line indicates the CDMS bound, 
the purple line the Xenon bound, 
and the green line the up-coming super CDMS bound.
The allowed region by LEP2 data is denoted as orange region.
} 
\label{fig:random120_j}
}
\FIGURE{
\epsfig{figure=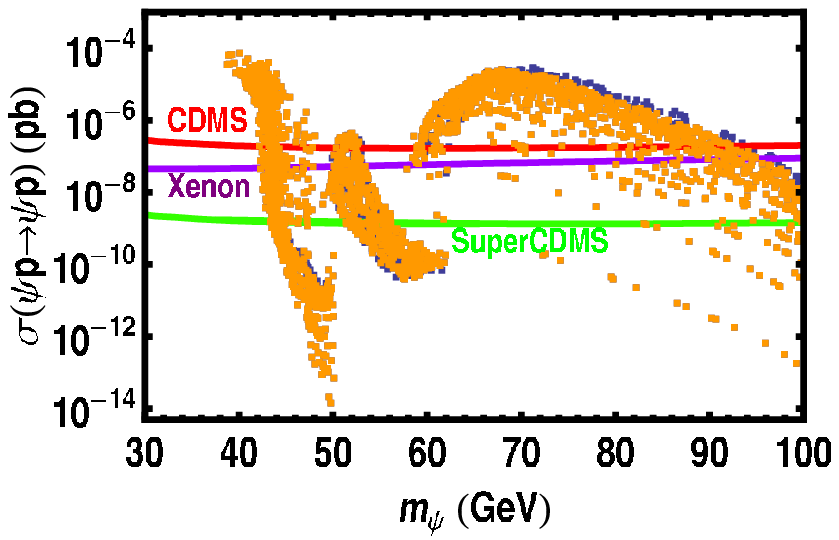,width=10cm,height=8cm}
\caption{
Predictions of the elastic scattering cross section 
$\sigma(\psi p \to \psi p)$ with respect to $m_{\psi}$ 
with $m_{h_1}=120$ GeV $(\pm 4\%)$ and $m_{h_2}=100$ GeV $(\pm 1\%)$. 
The red line indicates the CDMS bound, 
the purple line the Xenon bound, 
and the green line the up-coming super CDMS bound.
The allowed region by LEP2 data is denoted as orange region.
} 
\label{fig:random_mix}
}
Since $f^{(p,n)}_{Ts}$ dominantly contributes to $f_{(p,n)}$, 
we let $f_{p} \approx f_{n}$. 
The effective coupling constant $\alpha_q$ is defined 
by the spin-independent four fermion interaction 
of the quarks and the dark matter fermion in this model.
The effective Lagrangian is given by
\be
\mathcal{L}_{int} = \sum_{q} \alpha_q (\bar{\psi}\psi)(\bar{q}q) ,
\ee
where $\alpha_q$ is derived by the Higgs exchange $t$-channel diagram 
to be determined by
\be
\alpha_q = \frac{g_S \sin\theta \cos\theta m_q}{v_0}
           \left( \frac{1}{m_{h_1}^2}-\frac{1}{m_{h_2}^2} \right).
\ee
The elastic scattering cross section is obtained 
from the effective Lagrangian (\ref{effl})
\be
\sigma = \frac{4 M_r^2}{\pi}\left[Z f_p + (A-Z) f_n\right]^2
      \approx \frac{4 M_r^2 A^2}{\pi} f_p^2 ,
\ee
where $M_r$ is the reduced mass 
defined by $1/M_r = 1/m_{\psi} + 1/m_{nuclei}$. 
For the convenience of being compared with the experiments,
we obtain the cross section with the single nucleon given by
\be
\sigma(\psi p \to \psi p) \approx \frac{4 m_r^2}{\pi} f_p^2 ,
\ee
where $m_r$ is the reduced mass given by $1/m_r = 1/m_{\psi}+1/m_p$.

The prediction of the elastic scattering cross sections 
with the allowed parameter set as in Fig.\ref{fig:new90_c}
is depicted in Fig.\ref{fig:new90_j}. 
Allowed parameter sets by the LEP2 data
are again denoted by orange points on the figure.
The allowed region is entirely excluded by the current experiments 
including CDMS \cite{CDMS} and Xenon\cite{xenon}. 

The cross sections with the allowed parameter set 
of Fig.\ref{fig:random120_c} and Fig.\ref{fig:random_mix_c} 
are also depicted in 
Fig.\ref{fig:random120_j} and Fig.\ref{fig:random_mix}, respectively. 
The resonance region represented by a valley is not excluded 
by the experiments up to date. The large $m_\psi$ region in Fig.\ref{fig:random_mix}
is also still allowed.
We expect that the super CDMS experiment in the future
will probe the most region of the parameter set
for the singlet fermionic dark matter. 

\section{Conclusion}
\label{sec:Conclusion} 

We propose a renormalizable model with a fermionic cold dark matter. 
A minimal hidden sector consisting of a SM gauge singlet Dirac
fermion and a real singlet scalar is introduced.
We show that the singlet fermion can be a candidate
of the cold dark matter which explain
the relic abundance measured by WMAP.
The constraints on the masses and couplings at LEP2
are included and the elastic scattering cross sections
for the direct detection are predicted.
We find that most region of the parameter set
will be probed by the direct detection through elastic scatterings 
of the DM with nuclei in the near future.

\begin{acknowledgments}
This work was supported by 
the KRF Grant funded by the Korean Government (KRF-2005-C00006),
the KOSEF Grant (KOSEF R01-2005-000-10404-0), and the Center
for High Energy Physics of Kyungpook National University,
the BK21 program of Ministry of Education (Y.G.K., S.S.),
and the Korea Research Foundation Grant
funded by the Korean Government 
(MOEHRD, Basic Research Promotion Fund KRF-2007-C00145)
and the BK21 program of Ministry of Education (K.Y.L.).
S.S. also thanks to Dr. Junghee Kim for useful discussions 
on the numerical work.
\end{acknowledgments}

\def\PRD #1 #2 #3 {Phys. Rev. D {\bf#1},\ #2 (#3)}
\def\PRL #1 #2 #3 {Phys. Rev. Lett. {\bf#1},\ #2 (#3)}
\def\PLB #1 #2 #3 {Phys. Lett. B {\bf#1},\ #2 (#3)}
\def\NPB #1 #2 #3 {Nucl. Phys. {\bf B#1},\ #2 (#3)}
\def\ZPC #1 #2 #3 {Z. Phys. C {\bf#1},\ #2 (#3)}
\def\EPJ #1 #2 #3 {Euro. Phys. J. C {\bf#1},\ #2 (#3)}
\def\JHEP #1 #2 #3 {JHEP {\bf#1},\ #2 (#3)}
\def\JCAP #1 #2 #3 {JCAP {\bf#1},\ #2 (#3)}
\def\JPG #1 #2 #3 {J. of Phys. G {\bf#1},\ #2 (#3)}
\def\IJMP #1 #2 #3 {Int. J. Mod. Phys. A {\bf#1},\ #2 (#3)}
\def\MPL #1 #2 #3 {Mod. Phys. Lett. A {\bf#1},\ #2 (#3)}
\def\PTP #1 #2 #3 {Prog. Theor. Phys. {\bf#1},\ #2 (#3)}
\def\PR #1 #2 #3 {Phys. Rep. {\bf#1},\ #2 (#3)}
\def\RMP #1 #2 #3 {Rev. Mod. Phys. {\bf#1},\ #2 (#3)}
\def\PRold #1 #2 #3 {Phys. Rev. {\bf#1},\ #2 (#3)}
\def\IBID #1 #2 #3 {{\it ibid.} {\bf#1},\ #2 (#3)}

\end{document}